# Quantum no-key protocol for direct and secure transmission of quantum and classical messages

Li Yang[*]

State Key Laboratory of Information Security (Graduate School of Chinese Academy of Sciences), Beijing 100039, P. R. China

Various protocols of quantum cryptography has been presented [1-4]. Some of them can be used to transmit secret messages directly[5-7]. In this letter, we present a new kind of quantum cryptography protocol for direct transmission of classical and quantum messages based on Shamir's generic protocol on classical message encryption [8]. Though Shamir's original idea has only realization of computational security in classical cryptology and will fall to a man-in-the-middle attack, our protocol's security is theoretical based on properties of quantum entanglement and Boolean function, and can resist the man-in-the-middle attack.

## I. Protocol for quantum message transmission

A quantum message is a sequence of pure state:

$$M_k^{(n)} = \{\sum_m \alpha_m^{(i)} |m\rangle \,|\, i = 1, 2, \cdots, n\}, \tag{1}$$

where $m = (m_1, m_2, \cdots, m_k) \in F_2^{(k)}$ if we choose $|m_1\rangle \otimes |m_2\rangle \otimes \cdots \otimes |m_k\rangle$ as base states. Let us consider the secure transmission of a pure state $\sum_m \alpha_m |m\rangle$. Here the word 'secure' means 1) Eve cannot get the state even when she has controlled the channel; 2) Bob can verify that the state really comes from Alice; 3) Alice can verify that the receiver is Bob; 4) Bob know whether the state has been changed in the channel. These are so called encryption, identification and authentication. Here is the basic encryption protocol for quantum message without authentication:

1. Alice chooses a n-dimensional Boolean function

$$F_A(x) = \left(F_A^{(1)}(x), F_A^{(2)}(x), \cdots, F_A^{(n)}(x)\right) \tag{2}$$

from Boolean functions of k variables randomly and secretly, and do a computation as below:

$$U_{F_A} \sum_m \alpha_m |m\rangle_I |0\rangle_{II} = \sum_m \alpha_m |m\rangle_I |F_A(m)\rangle_{II}, \tag{3}$$

then sends the state to Bob.

2. Bob chooses his Boolean function $F_B(x)$ independently and randomly, and do a similar computation:

$$U_{F_B} \sum_m \alpha_m |m\rangle_I |F_A(m)\rangle_{II} |0\rangle_{III} = \sum_m \alpha_m |m\rangle_I |F_A(m)\rangle_{II} |F_B(m)\rangle_{III}, \tag{4}$$

then sends it back to Alice.

---
[*] E-mail address: yangli@gscas.ac.cn

3. Alice do the computation:

$$U_{F_B} \sum_m \alpha_m |m\rangle_I |F_A(m)\rangle_{II} |F_B(m)\rangle_{III}$$
$$= \sum_m \alpha_m |m\rangle_I |F_A(m) \oplus F_A(m)\rangle_{II} |F_B(m)\rangle_{III} = \sum_m \alpha_m |m\rangle_I |0\rangle_{II} |F_B(m)\rangle_{III} \quad (5)$$

and sends

$$\sum_m \alpha_m |m\rangle_I |F_B(m)\rangle_{III} \quad (6)$$

to Bob again.

4. Bob do the same computation with his function $F_B(x)$:

$$\sum_m \alpha_m |m\rangle_I |F_B(m)\rangle_{III} \to \sum_m \alpha_m |m\rangle_I |F_B(m) \oplus F_B(m)\rangle_{III}$$
$$= \sum_m \alpha_m |m\rangle_I |0\rangle_{III} \quad (7)$$

and get the quantum message in the first quantum register.

## II. Protocol for classical message transmission

Alice prepares the base state $|m'\rangle$ in a quantum register of k qubits to represents a classical message $m'$ of k bits, then transforms it to a superposition state via Hadamard transformation:

$$U_H |m_1', m_2', \cdots, m_k'\rangle = \frac{1}{\sqrt{2^k}} \sum_{m_1, m_2, \cdots, m_k} (-1)^{m_1 m_1' + m_2 m_2' + \cdots + m_k m_k'} |m_1, m_2, \cdots, m_k\rangle, \quad (8)$$

where $m_i'(m_i)$ is the value of the $i$ th bit of message $m'(m)$. After that, Alice transmits it with the protocol described in the previous section. In the end, Bob should transform the state he has received to a base state via Hadamard transformation to get the classical message $m'$.

Let us consider the simplest example for the classical message transmission: Alice want to transmit a single bit message '0' to Bob. The process is as below:

1. Alice chooses $|0\rangle$ to represents '0', and transforms it to a superposition state with Hadamard transformation:

$$|0\rangle \to \frac{1}{\sqrt{2}}(|0\rangle + |1\rangle),$$

then computes the Boolean function $F_A$:

$$U_{F_A}\left[\frac{1}{\sqrt{2}}(|0\rangle_I + |1\rangle_I)|0\rangle_{II}\right] = \frac{1}{\sqrt{2}}(|0\rangle_I |F_A(0)\rangle_{II} + |1\rangle_I |F_A(1)\rangle_{II}), \quad (9)$$

here $F_A$ is chosen from $F(x) = x, \bar{x}, 1, 0$ randomly. It is evident that the computation involved here can be realized by a CNOT gate and a single qubit gate. After the computation, Alice sends the two-qubit state to Bob.

2. Bob computes the Boolean function $F_B$ chosen by himself :

$$U_{F_B}\left[\frac{1}{\sqrt{2}}\big(|0\rangle_I|F_A(0)\rangle_{II}+|1\rangle_I|F_A(1)\rangle_{II}\big)|0\rangle_{III}\right]$$
$$=\frac{1}{\sqrt{2}}\big(|0\rangle_I|F_A(0)\rangle_{II}|F_B(0)\rangle_{III}+|1\rangle_I|F_A(1)\rangle_{II}|F_B(1)\rangle_{III}\big). \tag{10}$$

Then Bob sends this three-qubit state back to Alice.

3. Alice computes $F_A$ and gets

$$\frac{1}{\sqrt{2}}\big[|0\rangle_I|0\rangle_{II}|F_B(0)\rangle_{III}+|1\rangle_I|0\rangle_{II}|F_B(1)\rangle_{III}\big], \tag{11}$$

and then sends the state

$$\frac{1}{\sqrt{2}}\big[|0\rangle_I|F_B(0)\rangle_{III}+|1\rangle_I|F_B(1)\rangle_{III}\big] \tag{12}$$

back to Bob.

4. Bob computes $F_B$ again and gets

$$\frac{1}{\sqrt{2}}\big(|0\rangle_I+|1\rangle_I\big)|0\rangle_{III}, \tag{13}$$

Then he do a Hadamard transformation to the first qubit and get $|0\rangle$ state. Because the Boolean functions $F_A$ and $F_B$ are chosen randomly and secretly from $F=0,1,x,\bar{x}$, the protocol is theoretically secure.

## III. Protocol with personal identification and message authentication

Suppose Alice and Bob preshare identification keys $s_A$ and $s_B$, here $s_A$ and $s_B$ are Boolean functions. We have a protocol that will success against the 'middle-man attack' :

1. Alice prepares the state as below:

$$\sum_m \alpha_m|m\rangle_I|0\rangle_{II} \to \sum_m \alpha_m|m\rangle_I|F_A(m)\rangle_{II}, \tag{14}$$

and sends it to Bob.

2. Bob transforms it :

$$\to \sum_m \alpha_m|m\rangle_I|F_A(m)\oplus s_B(m)\rangle_{II}|F_B(m)\rangle_{III}, \tag{15}$$

and sends it back to Alice.

3. Alice transforms the state and verifies that the quantum message is really coming back from Bob:

$$\to \sum_m \alpha_m |m\rangle_I |F_A(m)\rangle_{II} |F_B(m)\rangle_{III} \to \sum_m \alpha_m |m\rangle_I |0\rangle_{II} |F_B(m) \oplus s_A(m)\rangle_{III}. \quad (16)$$

That is, if the second quantum register is in the state $|0\rangle$, Alice believe that it really comes from Bob, otherwise she stop the protocol, or more safely, process a wrong protocol. If Eve pretend to be Alice to communicate with Bob, she can substitute the second register with one in the state $|F_E(m)\rangle$, but she cannot transform the third register into $|F_B(m) \oplus s_A(m)\rangle$ if we choose $s_A \neq s_B$. Finally Alice transforms the state to

$$\sum_m \alpha_m |m\rangle_I |F_B(m) \oplus s_A(m)\rangle_{III}, \quad (17)$$

and sends it to Bob again.

4. Bob transforms the state as below to get the quantum message coming form Alice:

$$\to \sum_m \alpha_m |m\rangle_I |F_B(m)\rangle_{III} \to \sum_m \alpha_m |m\rangle_I |0\rangle_{III}, \quad (18)$$

and at the same time, to verify Alice's legitimacy via measuring the third register.

## IV. Discussions and conclusion

1. We can find that there are at least other two choices to construct no-key like protocol for pure state transmission. The simplest one is Alice and Bob changes the bases directly and interactively:

$$\sum_m \alpha_m |m \oplus s_A\rangle \to \sum_m \alpha_m |m \oplus s_A \oplus s_B\rangle \to \sum_m \alpha_m |m \oplus s_B\rangle \to \sum_m \alpha_m |m\rangle. \quad (19)$$

Unfortunately, we cannot authenticate the state at all. The other choice is to make use an auxiliary register:

$$\sum_m \alpha_m |m\rangle |F_A(m)\rangle \to \sum_m \alpha_m |m\rangle |F_A(m) \oplus F_B(m)\rangle$$
$$\to \sum_m \alpha_m |m\rangle |F_B(m)\rangle \to \sum_m \alpha_m |m\rangle \quad , \quad (20)$$

but still we cannot construct a secure protocol with both encryption and authentication.

2. Suppose Alice and Bob share secret key s, they can use it directly as

$$\sum_m \alpha_m |m\rangle \to \sum_m \alpha_m |m \oplus s\rangle \to \sum_m \alpha_m |m\rangle$$

if s is a random string. If s is a Boolean function, they have at least two other choices:

$$\sum_m \alpha_m |m\rangle \to \sum_m \alpha_m |s(m)\rangle \to \sum_m \alpha_m |m\rangle, \quad (21)$$

and

$$\sum_m \alpha_m |m\rangle \to \sum_m \alpha_m |m\rangle |s(m)\rangle \to \sum_m \alpha_m |m\rangle. \quad (22)$$

It is easy to find that we have no way to authenticate the message at all. Besides, Eve can get information of s by the way of measuring the state in the channel. Our protocol uses $F_A$ and $F_B$ to protect Alice and Bob's personal keys $s_A$ and $s_B$.

## Referencies